\documentstyle[tighten,preprint,eqsecnum,aps,prd]{revtex}

\begin{document}
\draft

\preprint{\vbox{\baselineskip=12pt
\rightline{WISC--MILW--95--TH--19}
\rightline{PP96--36}
\rightline{gr-qc/9510048}}}
\title{Hamiltonian thermodynamics of
2D vacuum dilatonic \\ black holes}
\author{Sukanta Bose\footnote{Electronic address:
{\em bose@csd.uwm.edu}}}
\address{Department of Physics,
University of Wisconsin-Milwaukee, \\
P.~O. Box 413,
Milwaukee, Wisconsin 53201, USA}
\author{Jorma Louko\footnote{Electronic address:
{\em louko@wam.umd.edu\/}.
On leave of absence from
Department of Physics,
University of Helsinki.}}
\address{Department of Physics,
University of Maryland, \\
College Park, Maryland 20742-4111, USA}
\author{Leonard Parker\footnote{Electronic address:
{\em leonard@cosmos.phys.uwm.edu}}
and
Yoav Peleg\footnote{Electronic address:
{\em yoav@csd.uwm.edu}}
}
\address{Department of Physics,
University of Wisconsin-Milwaukee, \\
P.~O. Box 413,
 Milwaukee, Wisconsin 53201, USA}
\date{Final version; published in Physical Review D 53, 1996, pp. 5708-5716}

\maketitle
\begin{abstract}%
We consider the Hamiltonian dynamics and thermodynamics of the
two-dimensional vacuum dilatonic black hole in the presence of a timelike
boundary with a fixed value of the dilaton field. A~canonical transformation,
previously developed by Varadarajan and Lau, allows a reduction of the
classical dynamics into an  unconstrained Hamiltonian system with one
canonical pair of degrees of freedom. The reduced theory is quantized, and
a partition function of a canonical ensemble is obtained as the trace of
the analytically continued time evolution operator. The partition function
exists for any values of the dilaton field and the temperature at the
boundary, and the heat capacity is always positive. For temperatures
higher than $\beta_c^{-1} = \hbar\lambda/(2\pi)$, the partition function
is dominated by a classical black hole solution, and the dominant
contribution to the entropy is the two-dimensional Bekenstein-Hawking
entropy. For temperatures lower than~$\beta_c^{-1}$, the partition function
remains well-behaved and the heat capacity is positive in the asymptotically
flat space limit, in contrast to the corresponding limit in
four-dimensional spherically symmetric Einstein gravity; however, in this
limit, the partition function is not dominated by a classical black hole
solution.
\end{abstract}

\narrowtext

\section{Introduction}
\label{sec:intro}

The observation that two-dimensional (2D)
dilaton gravity theories admit classical
and semiclassical black hole solutions\cite{Witten,CGHS} has inspired an
intense interest in these theories as an arena for studying quantum black
holes. At the very least, these theories provide a simplified arena for the
study of the final stages of black hole evaporation, in the hope that the
two-dimensional results would in some broad sense reflect on
four-dimensional (4D) quantum gravity. More ambitiously, one may entertain
the hope that two-dimensional dilaton gravity could in fact be closer to the
`true' theory of gravity in our universe than four-dimensional Einstein
gravity. For a review, see Ref.\cite{stro-rev}.

The purpose of the present paper is to investigate the equilibrium
thermodynamics of the two-dimensional vacuum dilatonic black
hole\cite{Witten,CGHS} (``Witten's black hole") in the canonical ensemble.
Within four-dimensional Einstein gravity, it is well known that the
canonical ensemble for the radiating Schwarzschild black hole does not exist
in asymptotically flat space\cite{GH1,hawkingCC}, but the situation can be
improved by postulating the black hole to be placed in a spherical,
mechanically rigid `box' on which the local temperature is then
held fixed\cite{york1,WYprl,whitingCQG}; see Ref.\cite{pagerev} for a review.
This motivates us to consider the canonical ensemble for the dilaton black
hole under analogous boundary conditions, with the dilaton field providing
the analogue of the box radius. As a limiting case, we shall also be able to
address the canonical ensemble in asymptotically flat space.

For four-dimensional spherically symmetric Einstein gravity with a finite
boundary, the evaluation of the thermodynamical partition function in the
canonical ensemble has been addressed through a combination of Hamiltonian
and path integral
techniques\cite{york1,WYprl,whitingCQG,LWo,BY-quasilocal,BY-microcan,MW,LW}.
Also, in two-dimensional dilaton gravity, a similar calculation of
the thermodynamical partition function using the path
integral approach exists \cite{Solodukhin}.
In this paper we shall adapt to the dilaton gravity theory the Hamiltonian
method of Ref.\cite{LW} (henceforth referred to as LW). 

In this method, one
first constructs a classical Lorentzian Hamiltonian theory of geometries such
that, on the classical solutions, one end of the spacelike surfaces is on a
timelike boundary in an exterior region of the black hole spacetime, and the
other end is at the horizon bifurcation surface. One then canonically
quantizes this theory, and obtains the thermodynamical partition function by
suitably continuing the Schr\"odinger picture time evolution operator to
imaginary time and taking the trace. A~crucial input is how to do the
analytic continuation at the bifurcation surface; in LW it was found that a
choice motivated by smoothness of Euclidean black hole geometries yields a
partition function that is in agreement with that obtained via path integral
methods.

In order that the method can be implemented, one must be able to
canonically quantize the Lorentzian theory in some practical fashion.
In~LW this was achieved for spherically symmetric four-dimensional
Einstein gravity by using canonical variables that were first
introduced by Kucha\v{r} under asymptotically flat, Kruskal-like boundary
conditions\cite{KVK}. In these variables the constraints become
exceedingly simple, and the classical Hamiltonian theory could be
explicitly reduced into an unconstrained Hamiltonian theory with just one
canonical pair of degrees of freedom. For the two-dimensional dilaton gravity
theory,  an analogue of Kucha\v{r}'s variables was recently found by
Varadarajan \cite{Madhavan} under Kruskal-like boundary conditions, and
by Lau \cite{Lau} under boundary conditions analogous to those
in~LW\null.\footnote{Ref.\cite{Lau} appeared while the present paper was in
preparation.} We shall see that using these variables, it will be possible to
construct a quantum theory and a thermodynamical partition function for the
dilatonic theory in close analogy with~LW\null.

As in spherically symmetric Einstein gravity in four dimensions, the
partition function in the dilatonic theory will turn out to exist for all
values of the dilaton and the temperature at the boundary. The heat capacity
is always positive, implying thermodynamical stability of the canonical
ensemble. For temperatures higher than the critical value $\beta_c^{-1} =
\hbar \lambda/(2\pi)$, the partition function is dominated by a classical
black hole solution, and the dominant contribution to the entropy is simply
the two-dimensional Bekenstein-Hawking entropy~$S_{BH}$,
\begin{equation}
\label{sclass}
S_{BH}=\beta_c M
= {2\pi M \over \hbar \lambda},
\end{equation}
where $M$ is the ADM mass \cite{Witten,Madhavan,bilal} of the hole. For
temperatures lower than~$\beta_c^{-1}$, on the other hand, the partition
function is not dominated by a classical black hole solution. These
properties are easily understood physically in terms of the gravitational
blueshift effect and the fact that $\beta_c^{-1}$ is the Hawking temperature
at infinity for the dilatonic black hole with any value of the
mass\cite{Witten}. The main difference from four-dimensional spherically
symmetric Einstein gravity is that in the four-dimensional case, the
condition that the entropy be dominated by the Bekenstein-Hawking entropy of
a classical black hole solution is that the product of the temperature and
the boundary curvature radius be larger than a critical numerical
value\cite{york1,WYprl}. Also, a four-dimensional black hole solution that
dominates the partition function is necessarily so massive that the box is
contained within the $3M$ radius\cite{york1}, and for the
partition function of Refs.\cite{WYprl,whitingCQG} even inside
the $\case{9}{4}M$ radius; in contrast, for a two-dimensional black hole
solution that dominates the partition function, the box can be arbitrarily
large compared with the length scale set by the mass.

For temperatures lower than~$\beta_c^{-1}$, taking the boundary
to infinity yields a well-defined partition function, which can be
identified with the partition function associated with asymptotically
flat boundary conditions at infinity. This is in a striking contrast
with four-dimensional spherically symmetric Einstein theory, where
the partition function diverges in the asymptotically flat space
limit\cite{york1,WYprl}. In the two-dimensional case,
the heat capacity turns out to be again positive,
but it diverges as the temperature approaches the critical
value~$\beta_c^{-1}$. Again, this behavior is easily understandable in view
of the classical black hole solutions. As the Hawking temperature at
infinity is independent of the ADM mass, the hole can absorb or emit energy
without changing its temperature: the heat capacity can thus be regarded as
infinite.

The rest of the paper is as follows. In Section
\ref{sec:dilaton-theory} we briefly recall the two-dimensional vacuum
dilaton gravity theory and Witten's black hole solution, establishing our
notation which is motivated by the reduction from four-dimensional dilaton
gravity\cite{Horowitz,Peleg}. In Section \ref{sec:geometrodynamics} we
present, in these variables, a canonical transformation which is equivalent
to that given by Lau\cite{Lau}, and differs from that given by
Varadarajan\cite{Madhavan} in essence only in the boundary conditions. We
also reduce the theory to a single true pair of canonical variables. The
partition function is constructed and the thermodynamical properties are
discussed in Section~\ref{sec:thermodynamics}, and the limit of
asymptotically flat space is explored in Section~\ref{sec:flat}.
Section~\ref{sec:conclusions} offers brief concluding remarks.

We shall work throughout in units in which $c=G=1$. The action will then be
dimensionless, which means that also the two-dimensional Planck's constant
$\hbar$ is dimensionless.

\section{2D dilaton gravity}
\label{sec:dilaton-theory}

\subsection{Dynamics}
\label{subsec:dynamics}

The CGHS action for a two-dimensional theory of gravity coupled to a
dilaton field $\phi$ is\cite{CGHS}
\begin{equation}
\label{action1}
S = \case{1}{2} \int dt \int dr \sqrt{-g}e^{-2\phi}
        [{\cal R}^{(2)}+4(\nabla \phi)^2+4\lambda^2],
\end{equation}
where ${\cal R}^{(2)}$ is the two-dimensional Ricci scalar and $\lambda^2$
is a cosmological constant term. We shall take $\lambda>0$.  The
two-dimensional metric can be written in the ADM form \cite{ADM}
\begin{equation}
\label{ADMmetric}
ds^2 =-N^2 dt^2 +\Lambda^2(dr+N^r dt)^2,
\end{equation}
where the lapse-function~$N$, the shift vector~$N^r$, and the
spatial one-metric component $\Lambda^2$ are functions of $t$ and~$r$. We
shall assume $\Lambda$ and $N$ to be positive, which guarantees that the
metric is nondegenerate with signature~$(-+)$. Viewing the two-dimensional
CGHS theory as a reduced four-dimensional dilatonic gravity
theory\cite{Horowitz}, we define the positive-definite field $R = e^{-2\phi}
/ (2\lambda)$, which can be interpreted as the radial coordinate for the
four-dimensional classical solution\cite{Peleg}. Up to
boundary terms, the action (\ref{action1}) takes then the ADM form
\begin{eqnarray}
\label{ADMaction}
S = \int dt \int dr \Bigglb\{
&& \left( -{2\lambda\over RN}\right)
  \left[R \left(-\dot{\Lambda}+(N^r\Lambda)'\right)
  (-\dot{R}+R'N^r)+ \case{1}{2} \Lambda (-\dot{R}+R'N^r)^2 \right]
\nonumber \\
  &&+ 2\lambda R' \Lambda^{-1}N'+\lambda\Lambda^{-1}NR^{-1}(R')^2
  +4\lambda^3NR\Lambda \Biggrb\} ,
\end{eqnarray}
where~$\>\dot{}\>={\partial\over \partial t}$
  and~$\> '\> ={\partial\over \partial r}$.

For concreteness, we take both $r$ and $t$ to have the dimension of length,
which implies that $\Lambda$, $N$, and $N^r$ are dimensionless. The constant
$\lambda$~has the dimension of inverse length, $R$~has the dimension of
length, and the action is dimensionless. We shall follow the convention of
Kucha\v{r}\cite{KVK} in denoting by Latin letters those canonical
coordinates that are spatial scalars (e.g.~$R$), and by Greek letters those
that are spatial densities (e.g.~$\Lambda$).

The momenta conjugate to $\Lambda$ and $R$ are respectively
\begin{mathletters}
\begin{eqnarray}
\label{piL}
\Pi_{\Lambda}
&=&
-2\lambda N^{-1}(\dot{R}-R'N^r),
\\
\label{piR}
\Pi_{R}
&=&
-2\lambda N^{-1}\left[ \Lambda R^{-1}(\dot{R}-R'N^r)
                          +\left(\dot{\Lambda}-(N^r\Lambda)' \right)
                          \right].
\end{eqnarray}
\end{mathletters}%
The dimension of $\Pi_{\Lambda}$ is inverse length, and that of
$\Pi_{R}$ is inverse length squared. Note that the momentum with a Greek
subscript is a spatial scalar and that with a Latin subscript is a
spatial density. A~Legendre transformation leads to the canonical bulk
action (i.e., the action up to boundary terms):
\begin{equation}  \label{canaction}
S_\Sigma [R,\Lambda,\Pi_R,\Pi_{\Lambda};N,N^r]
=
\int dt
\int dr \,
(\dot{R}\Pi_R  +\dot{\Lambda}\Pi_{\Lambda}-NH-N^r H_r),
\end{equation}
where the super-Hamiltonian $H$ and the supermomentum $H_r$ are given
respectively by
\begin{mathletters}
\begin{eqnarray} \label{superH}
H &=&
-{(2\lambda)}^{-1}\Pi_R \Pi_{\Lambda}
+{(4\lambda)}^{-1}R^{-1}\Lambda
  \Pi_{\Lambda}^2
\nonumber \\
&& +2\lambda \Lambda^{-1}R''-2\lambda \Lambda^{-2}R'
  \Lambda ' -\lambda \Lambda^{-1}R^{-1}{R'}^2
  -4\lambda^3 \Lambda R,
\\
H_r &=&
\Pi_R R'-\Lambda {\Pi_{\Lambda}}'.
\label{superHr}
\end{eqnarray}
\end{mathletters}%
We shall discuss the boundary conditions and boundary terms in
Section~\ref{sec:geometrodynamics}.

\subsection{Witten's black hole}

The general solution\cite{CGHS,Peleg} to the equations of motion
derived from the action (\ref{ADMaction}) is
\begin{equation}
\label{Rmetric}
ds^2=-F(R)dT^2 + {dR^2\over {4\lambda^2 R^2 F(R)}}\quad ,
\end{equation}
where
\begin{equation}
\label{F}
F(R)=1-{M\over {2\lambda^2 R}}
   \quad .
\end{equation}
Here $M$ is a parameter that can be interpreted as the ADM
mass\cite{Witten,Madhavan,bilal}. For positive~$M$, Eq.~(\ref{Rmetric})
describes a two-dimensional black hole geometry \cite{CGHS}, with a horizon
at $R=M/(2\lambda^2)$ and a spacelike singularity at $R=0$.
The $(T, R)$ coordinates are analogous to the curvature coordinates
(``Schwarzschild coordinates") of the four-dimensional Schwarzschild
metric\cite{KVK}, and they cover at a time only one quadrant of the full
spacetime. In the global, Kruskal-like coordinates $(x^+,x^-)$ defined
via\cite{Peleg}
\begin{mathletters}
\begin{eqnarray}
\lambda^2 x^+ x^- &=& {M\over \lambda}-2\lambda R, \\
\ln |x^+/x^-| &=& 2\lambda T,
\end{eqnarray}
\end{mathletters}%
the metric takes the more familiar form
\begin{equation}  \label{xmetric}
ds^2= {-dx^+ dx^-\over {-\lambda^2 x^+ x^- +M/\lambda}}.
\end{equation}

Given the canonical data $(R,\Lambda,\Pi_R,\Pi_\Lambda)$ on one
hypersurface in the solution~(\ref{Rmetric}), one can read from this data
the value of the mass parameter $M$, and also the location of the surface up
to translations in the Killing time~$T$. Adapting Kucha\v{r}'s analysis of
the Schwarzschild black hole\cite{KVK} as in Refs.\cite{Madhavan,Lau}, one
finds
\begin{equation}
\label{Fcan} F=(2\lambda)^{-2}\left[ \left({R'\over {\Lambda R}}\right)^2
                  -\left({\Pi_{\Lambda}\over {2\lambda R}}\right)^2
\right]
\end{equation}
and
\begin{equation}
\label{Tprime}
-T'=(2\lambda)^{-2} R^{-1}F^{-1}\Lambda \Pi_{\Lambda}.
\end{equation}
$M$ is then recovered from (\ref{F}) using~(\ref{Fcan}).

It is now possible to follow Kucha\v{r}\cite{KVK} and promote the on-shell
expressions for $M$ and $T'$ into a canonical transformation, provided the
boundary conditions can be handled in a satisfactory manner. For Kruskal-like
boundary conditions this was achieved in Ref.\cite{Madhavan}, and for
thermodynamically motivated boundary conditions analogous to those of LW in
Ref.\cite{Lau}. In the following section we shall review this analysis under
the thermodynamical boundary conditions, in terms of the
variables introduced in subsection~\ref{subsec:dynamics}, and we shall
explicitly reduce the theory into its unconstrained Hamiltonian form.

\section{Geometrodynamics of Witten's black hole in a box}
\label{sec:geometrodynamics}

\subsection{Hamiltonian formulation}
\label{subsec:old-hamiltonian}

Our first task is to specify a set of boundary conditions analogous to those
of~LW, and to add to the Hamiltonian bulk action (\ref{canaction})
appropriate boundary terms. As the spatial proper distance will under our
boundary conditions be finite, we follow Refs.\cite{LW,Lau} and take $r$ to
have the range $[0,1]$.

Consider first the left end of the spacelike surfaces. At the limit
$r\rightarrow 0$, we adopt the fall-off conditions
\begin{mathletters}
\label{falloff}
\begin{eqnarray}
\Lambda (t,r) &=& \Lambda_0 (t) + O(r^2) , \label{lbp} \\
R(t,r) &=& R_0 (t) + R_2 (t) r^2 + O(r^4) , \label{rbp} \\
\Pi_{\Lambda} (t,r) &=& O(r^3) , \label{plbp} \\
\Pi_R (t,r) &=& O(r) , \label{prbp} \\
N(t,r) &=& N_1 (t) r + O(r^3) , \label{nbp} \\
N^r (t,r) &=& N^r_1 (t) r + O(r^3) , \label{nrbp}
\end{eqnarray}
\end{mathletters}%
where $\Lambda_0$ and $R_0$ are positive, and $N_1 \geq 0$. The consistency
of these conditions with the equations of motion can be shown as in
LW\cite{Lau}. When the equations of motion hold, the conditions enforce
$r=0$ to be at the horizon bifurcation point of a black hole solution.

Depending on what boundary conditions one chooses to impose on the
canonical data, the bulk action in (\ref{canaction}) needs to be
supplemented with boundary terms such 
that the variation of the total
action under the chosen boundary conditions leaves only a bulk
term that gives the equations of motion.
Consider now the total action $S=S_{\Sigma} + S_{\partial \Sigma}$, where
the bulk action $S_{\Sigma}$ was given in (\ref{canaction}) and the
boundary action $S_{\partial \Sigma}$ is given by
\begin{eqnarray}
\label{sbdry}
S_{\partial \Sigma}[R,\Lambda,\Pi_R,\Pi_{\Lambda};N,N^r]
&=&2\lambda \int dt [RN' \Lambda^{-1}]_{r=0} \nonumber \\
&&+\int dt \left[ 2\lambda N R' \Lambda^{-1}-N^r \Lambda \Pi_{\Lambda}
-\lambda \dot{R} \ln \left| {N+\Lambda N^r \over {N-\Lambda N^r}}
                    \right| \right]_{r=1} .
\end{eqnarray}
The variation of the total action contains a bulk term that yields the
equations of motion, as well as several boundary terms. The boundary terms
on the initial and final surfaces have the usual form
$\pm \int_0^1 dr ( \Pi_{\Lambda} \delta \Lambda +
\Pi_R \delta R)$,
and they vanish provided one fixes the one-metric and the dilaton field
on the initial and final surfaces. The boundary terms at $r=0$ take the
form
\begin{equation}
\label{bifurpt}
2\lambda \int dt  \bigl[ R\delta ( N' \Lambda^{-1}) \bigr]_{r=0}
= 2\lambda \int dt R_0 \delta ( N_1 \Lambda_0^{-1}),
\end{equation}
which vanishes if we set $\delta ( N_1 \Lambda_0^{-1}) = 0$. As in the
four-dimensional case of~LW, fixing the quantity $N_1 \Lambda_0^{-1}$
means fixing in the classical solution the rate at which the unit normal
to the constant $t$ surface is boosted at the coordinate singularity at
the bifurcation point. The boundary term from $r=1$ is cumbersome, but it
can be verified to vanish for the classical solutions provided one fixes
the timelike one-metric component $g_{tt}=-N^2 + (\Lambda N^r)^2$ and~$R$.

Thus, the total action $S_{\Sigma}+S_{\partial \Sigma}$ is appropriate for
a variational principle that fixes the one-metric and the dilaton field on
the initial and final surfaces and also on the timelike boundary $r=1$, and
in addition the quantity $N_1 \Lambda_0^{-1}$ at $r=0$. As will be seen in
Section~\ref{sec:thermodynamics}, these boundary conditions are tailored
in view of the thermodynamics of Witten's black hole in a box. Fixing the
one-metric on the timelike boundary will translate into fixing the
temperature at the box that encloses the black hole, and fixing the value
of the dilaton field at the timelike boundary will specify the ``radius"
of this box: these conditions will lead into the thermodynamical 
canonical ensemble.
Fixing $N_1 \Lambda_0^{-1}$ at the bifurcation point will turn
out to yield the black hole entropy, in a way that can be related to the
regularity of the Euclidean black hole solutions.

Two remarks are in order. Firstly, although we have here found it
convenient to introduce the boundary conditions and boundary terms
intrinsically within the Hamiltonian theory, it would of course be possible
to translate the conditions into the Lagrangian theory and introduce
corresponding boundary terms to be added to the (1+1) split Lagrangian
action (\ref{ADMaction}) or to the covariant CGHS action~(\ref{action1}).
As discussed in Ref.\cite{Lau}, the boundary terms to be added to the CGHS
action would consist of the contributions
$\pm\case{1}{2}\int dx^a \, \sqrt{\mp {}^{(1)}g} \, e^{-2\phi}K$ from the
spacelike and timelike boundaries, where ${}^{(1)}g$ is the one-metric
component, $K$ is the extrinsic curvature, and $x^a$ is respectively
$r$ or~$t$, as well as additional
contributions from the bifurcation point and from the corners where the
timelike boundary meets the spacelike boundaries. We shall, however, not
need the explicit form of the Lagrangian actions here.

Secondly, given our boundary conditions, the choice of the boundary
action is not unique. For example, it would be possible to replace
$S_{\partial \Sigma}$ (\ref{sbdry}) by any expression that is equivalent
when the classical equations of motion hold: as the bulk term in the
variation enforces the classical equations of motion for $0<r<1$,
continuity of the variables implies that such a replacement does not
change the critical points of the total action. Also, as discussed
in~LW, it would be possible to leave $N_1\Lambda_0^{-1}$ free at $r=0$ if
the first term in (\ref{sbdry}) were replaced by $2\lambda
\int dt \, {\tilde N}_0 R_0$, where ${\tilde N}_0(t)$ is a new quantity
that is fixed in the variational principle: the stationarity of the new
action gives the equation of motion $N_1\Lambda_0^{-1}={\tilde N}_0$, and
the boundary data for the classical solutions remains exactly the same as
before. Modifications of this kind would not affect the reduced
Hamiltonian theory that we shall arrrive at in
Subsection~\ref{subsec:reduction}, or the thermodynamical analysis of
Section~\ref{sec:thermodynamics}. For concreteness, we shall adhere to the
boundary term $S_{\partial\Sigma}$~(\ref{sbdry}).

\subsection{Canonical transformation}
\label{subsec:new-hamiltonian}

We pass from the old canonical variables $(R,\Lambda,\Pi_R,\Pi_\Lambda)$
to the new canonical set $(M,R,P_M,P_R)$ via the equations
\begin{mathletters}
\begin{eqnarray}
\label{M}
M &=& {(8\lambda^2)}^{-1}
R^{-1}{\Pi^2}_{\Lambda}-{1\over2}
\Lambda^{-2}R^{-1}{R'}^2+2\lambda^2 R,
\\
\label{PM}
P_{M} &=& (2\lambda)^{-2} R^{-1}F^{-1}\Lambda \Pi_{\Lambda},
\\
P_R &=& \Pi_R - {1\over 2} R^{-1}\Lambda \Pi_{\Lambda}
      - {1\over 2} R^{-1}F^{-1}\Lambda \Pi_{\Lambda}
\nonumber
\\
&& - (2\lambda R)^{-2}F^{-1}\Lambda^{-2} \left[
      R'(\Lambda \Pi_{\Lambda})'-R''(\Lambda \Pi_{\Lambda}) \right],
\label{PR}
\end{eqnarray}
\end{mathletters}%
where $F$ is understood to be defined by~(\ref{Fcan}).
On a classical solution, the variable $M$ is a constant whose value is the
ADM mass parameter in the metric~(\ref{Rmetric}). Note that the variable
$R$ appears both in the old and new canonical sets, but the
transformation changes its conjugate momentum.

To prove that the transformation is canonical, one starts from the
identity
\begin{eqnarray}
\Pi_{\Lambda} \delta \Lambda + \Pi_R \delta R
-P_M \delta M - P_R \delta R
&=& \delta \left(\Lambda\Pi_{\Lambda}+\lambda R'
\ln \left|
{2\lambda R'-\Lambda\Pi_{\Lambda}\over
{2\lambda R'+\Lambda\Pi_{\Lambda}}}
\right| \right) \nonumber\\
&&-\left(\lambda \delta R \ln \left|
{2\lambda R'-\Lambda\Pi_{\Lambda}\over
{2\lambda R'+\Lambda\Pi_{\Lambda}}}
\right| \right)' ,
\label{integrands}
\end{eqnarray}
and integrates both sides with respect to $r$ from $r=0$ to $r=1$. The
contribution from the second term on the right hand side vanishes, and we
obtain
\begin{equation}
\label{Liouville}
\int_0^1 dr (\Pi_{\Lambda} \delta \Lambda + \Pi_R \delta R)
-\int_0^1 dr (P_M \delta M + P_R \delta R)
= \delta \omega,
\end{equation}
where
\begin{equation}  \label{omega}
\omega [\Lambda, \Pi_{\Lambda}, R]
=\int_0^1 dr \left(\Lambda\Pi_{\Lambda}+\lambda R' \ln \left|
{2\lambda R'-\Lambda\Pi_{\Lambda}\over {2\lambda R'+\Lambda\Pi_{\Lambda}}}
\right| \right).
\end{equation}
The functional (\ref{omega}) is well-defined, and the difference of the old
and new Liouville forms is thus an exact form. This shows that the
transformation is canonical.

The constraint terms $NH + N^r H_r$ in the old surface action
(\ref{canaction}) take the form $N^M M' + N^R P_R$, where
\begin{mathletters}
\begin{eqnarray}
\label{NM} N^M &=&
(2\lambda)^{-2} N^r R^{-1}F^{-1} \Lambda \Pi_\Lambda
- {(2\lambda R)}^{-1} N F^{-1}\Lambda^{-1} R' ,
\\
\label{NR} N^R &=& R' N^r -  {(2\lambda)}^{-1} \Pi_\Lambda N.
\end{eqnarray}
\end{mathletters}%
We can thus write the new surface action as
\begin{equation}  \label{newac}
S_{\Sigma}[M,R,P_M,P_R;N^M,N^R]=\int dt \int_0^1 dr
  (P_M \dot{M} + P_R \dot{R} - N^M M' - N^R P_R),
\end{equation}
where the quantities to be varied independently are $M$, $R$, $P_M$, $P_R$,
$N^M$, and~$N^R$.

We take the total action to be
\begin{equation}  \label{newtac}
S[M,R,P_M,P_R;N^M,N^R] = S_{\Sigma}[M,R,P_M,P_R;N^M,N^R]
                 + S_{\partial \Sigma}[M,R,P_M,P_R;N^M,N^R],
\end{equation}
where
\begin{eqnarray}
S_{\partial\Sigma}[M,R,P_M,P_R;N^M,N^R]
&=&-\int dt [MN^M]_{r=0} \nonumber \\
&&+ \int dt \Bigglb[ 4\lambda^2 R
  \sqrt{{F} Q^2 + {(2\lambda R)}^{-2}{\dot{R}}^2} \nonumber \\
&&+ \lambda \dot{R} \ln \left( { \sqrt{{F} Q^2 + {(2\lambda
R)}^{-2}{\dot{R}}^2}-{(2\lambda R)}^{-1} \dot{R} \over
{ \sqrt{{F} Q^2 + {(2\lambda
R)}^{-2}{\dot{R}}^2} + {(2\lambda R)}^{-1} \dot{R}}}\right)
\Biggrb]_{r=1}.
\label{ntac}
\end{eqnarray}
Here $F$ is defined by (\ref{F}) as before, and $Q^2$ is defined by
\begin{equation}
Q^2 =
-g_{tt} = {F} (N^M)^2 -{(2\lambda
R)}^{-2} {F}^{-1} (N^R)^2.
\label{QSQ}
\end{equation}
$Q^2$~need not have a definite sign for all values of~$r$, but at $r=1$
it is positive by virtue of our boundary conditions.

The variation of (\ref{newtac}) contains a bulk term proportional to the
equations of motion, and several boundary terms. The initial and final
surfaces contribute
$\pm \int_0^1 dr (P_M \delta M + P_R \delta R)$,
which vanish if we fix $M$ and $R$ on these surfaces. The boundary
term from $r=0$ is
\begin{equation}
\label{NMro}
-\int dt [M \delta N^M]_{r=0} = -\int dt R_0 \delta N_0^M.
\end{equation}
which vanishes provided we fix $N_0^M = \lim_{r\to0}N^M$. As the fall-off
conditions (\ref{falloff}) imply
\begin{equation}
\label{nom}
\Lambda_0^{-1}N_1 =- \lambda N_0^M,
\end{equation}
fixing $N_0^M$ has the interpretation mentioned in the lines following
(\ref{bifurpt}) in terms of the unit normal to the constant $t$ surfaces
at $r\to0$. Finally, the boundary terms from $r=1$ are cumbersome, but it
can be verified as in LW that they vanish on the classical solutions
provided one fixes $R$ and~$Q^2$.

Our final action (\ref{newtac}) agrees with that obtained by Lau\cite{Lau}:
it can be recovered from Eqs.\ (5.8) and (4.36) in Ref.\cite{Lau} by
setting $\alpha =2$, $y=2$, $\Psi = -(1/2)\ln (2\lambda R)$, and $\bar{N} =
Q$.

\subsection{Hamiltonian reduction}
\label{subsec:reduction}

Reduction of the action (\ref{newtac}) by solving the constraints
proceeds as in the four-dimensional case of~LW. The constraint $P_R =0$
implies that $R$ and $P_R$ drop out altogether. The constraint $M'=0$
implies
\begin{equation}
\label{bfm}
M(t, r) = {\bf m} (t).
\end{equation}
Substituting (\ref{bfm}) into (\ref{newtac}) gives the action
\begin{equation}
\label{redac}
S[{\bf m}, {\bf p}; N_0^M;R_B,Q_B] =
 \int dt ( {\bf p} \dot{\bf m} - {\bf h}),
\end{equation}
where
\begin{equation}
\label{bfp}
{\bf p} = \int_0^1 dr P_M.
\end{equation}
The reduced Hamiltonian  ${\bf h}$ is given by
\begin{equation}
\label{redh}
 {\bf h} = {\bf h}_H +  {\bf h}_B
\end{equation}
where
\begin{mathletters}
\begin{eqnarray}
\label{hH}
{\bf h}_H &=& N_0^M {\bf m},
\\
{\bf h}_B &=& - 4\lambda^2 R_B
\sqrt{{F}_BQ_B^2 + {(2\lambda R_B)}^{-2}{\dot{R}}_B^2} \nonumber \\
&&- \lambda {\dot{R}}_B \ln \left( { \sqrt{{F}_BQ_B^2 + {(2\lambda
R_B)}^{-2}{\dot{R}}_B^2}-{(2\lambda R_B)}^{-1} {\dot{R}}_B \over
{ \sqrt{{F}_BQ_B^2 + {(2\lambda R_B)}^{-2}{\dot{R}}_B^2} +
{(2\lambda R_B)}^{-1} {\dot{R}}_B}} \right).
\label{hB}
\end{eqnarray}
\end{mathletters}%
Here $R_B$ and $Q_B^2$ stand for the values of $R$ and $Q^2$ at $r=1$, and
${F}_B = 1-{\bf  m}{(2\lambda^2 R_B)}^{-1}$. $R_B$, $Q_B^2$, and $N_0^M$ are
considered to be prescribed functions of~$t$, satisfying  $R_B >0$,
$Q_B^2 >0$, and $N_0^M \leq 0$.

The variational principle associated with the action (\ref{redac})
fixes the initial and final values of~${\bf m}$. The equation of motion
for ${\bf m}$ reads $\dot{\bf m}=0$, which reflects the fact that on
a classical solution ${\bf m}$ is equal to the mass of the black hole. The
equation of motion for ${\bf p}$ can be related via (\ref{Tprime}),
(\ref{PM}), and (\ref{bfp}) to the difference of the evolution rates
$\dot{T}$ of the Killing time $T$ at the two ends of the constant
$t$ surfaces.

\section{Hamiltonian thermodynamics}
\label{sec:thermodynamics}

\subsection{Quantum theory}
\label{subsec:quantization}

We now take the value of the dilaton field at the boundary to be
time-independent, ${\dot{R}}_B=0$. The action becomes
\begin{equation}
\label{classac}
S[{\bf m}, {\bf p}; N_0^M,Q_B;B]
=\int dt ( {\bf p} \dot{\bf m} - {\bf h}),
\end{equation}
where the Hamiltonian ${\bf h}$ is
\begin{equation}
\label{qh}
{\bf h}=\left(1-\sqrt{1-{\bf m}{\left(2\lambda^2 B\right)}^{-1}}
             \right)4\lambda^2 BQ_B +N^M_0 {\bf m}.
\end{equation}
Here $B$ denotes the positive time-independent value of~$R_B$, and $Q_B>0$
and $N^M_0 \le 0$ are prescribed functions of $t$ as before. To
obtain~(\ref{qh}), we have added to (\ref{redh}) the term $4\lambda^2
BQ_B$, which renormalizes the value of the Hamiltonian so that ${\bf
h}({\bf m} = 0) = 0$. As the added term is independent of the canonical
variables, this renormalization does not affect the equations of motion;
it is analogous to the addition of the $K_0$ term in four-dimensional
Einstein gravity\cite{GH1,hawkingCC,york1,WYprl}. The canonical momentum
${\bf p}$ takes all real values, but the range of the canonical coordinate
${\bf m}$ is $0<{\bf m}<2\lambda^2 B$.

In the quantum theory, we take the Hilbert space to be
${\sf H}= L^2([0,2\lambda^2 B];\mu)$,
that is, the space of square integrable functions of ${\bf m}$
with respect to the inner product
\begin{equation}
\label{inpro}
(\psi,\chi)=\int_0^{2\lambda^2 B} \mu d {\bf m} \,
\overline{\psi ({\bf m})} \chi ({\bf m}),
\end{equation}
where $\mu({\bf m};B)$ is a smooth positive weight function.
More specific assumptions about $\mu$ will be made in
subsection~\ref{subsec:thermodynamics}.

We take the Hamiltonian operator $\hat{\bf h}$ to act by pointwise
multiplication by the function ${\bf h}({\bf m})$~(\ref{qh}):
$(\hat{\bf h}\psi)({\bf m}) = {\bf h}({\bf m})\psi({\bf m})$. The
unitary time evolution operator is
\begin{equation}  \label{teo1}
\hat{K} (t_2; t_1) = \exp \left[ -i \hbar^{-1} \int_{t_1}^{t_2} dt' \,
\hat{\bf h}(t') \right] ,
\end{equation}
and it acts in ${\sf H}$ by pointwise multiplication by the function
\begin{equation}
\label{teo2}
K({\bf m};T_B;\Theta_H) = \exp \left[ -i \hbar^{-1}
\left(1-\sqrt{1-{\bf m}{\left(2\lambda^2 B\right)}^{-1}}
\right) 4\lambda^2 BT_B + i \hbar^{-1} \lambda^{-1} {\bf m}\Theta_H
\right],
\end{equation}
where
\begin{mathletters}
\begin{eqnarray}
\label{TB}
T_B &=& \int_{t_1}^{t_2} dt \, Q_B,
\\
\label{TH}
\Theta_H &=& -\lambda \int_{t_1}^{t_2} dt \, N^M_0.
\end{eqnarray}
\end{mathletters}%
This means that $T_B$ and $\Theta_H$ are two independent evolution
parameters. $T_B$~is the proper time elapsed at the timelike boundary, and
$\Theta_H$ is the boost parameter elapsed at the bifurcation point. $\hat{K}
(t_2; t_1)$ depends on $t_1$ and $t_2$ only through $T_B$ and~$\Theta_H$, and
we can write it as $\hat{K} (T_B,\Theta_H)$.

\subsection{Partition function}
\label{subsec:partition}

We wish to obtain the partition function by analytically continuing
$\hat{K} (T_B,\Theta_H)$ to imaginary time and taking the trace. Since $T_B$
is the Lorentzian proper time elapsed at the timelike boundary, we set it
equal to~$-i \hbar \beta$, and we interpret $\beta$ as the inverse
temperature at the boundary. The continuation of $\Theta_H$ is motivated by
consistency with the Euclidean path integral approach (i.e., requiring the
absence of a conical singularity in the Euclidean solution) as in~LW,
leading us to set $\Theta_H =-2\pi i$.
In this way we obtain for the partition
function the formal expression
\begin{eqnarray}
\label{partition1}
Z(\beta)&=&{\rm Tr} \left[\hat{K} (-i\hbar \beta,-2\pi i)\right]
\nonumber
\\
        &=& \int_0^{2\lambda^2 B} \mu d{\bf m} \,
\langle {\bf m} | \hat{K} ( -i\hbar\beta,-2\pi i ) |{\bf m} \rangle
\nonumber
\\
&=& \int_0^{2\lambda^2 B} \mu d{\bf m} \,
K ({\bf m};-i\hbar\beta ; -2\pi i)
\langle {\bf m}|{\bf m}\rangle,
\end{eqnarray}
where the last expression is divergent because of the inner product
$\langle {\bf m}|{\bf m}\rangle$.
The origin of this divergence can be traced to the absence of a kinetic
term in the Hamiltonian operator $\hat{\bf h}$ appearing in the
expression for the time evolution operator $\hat{K}$ given in~(\ref{teo1}).
To renormalize this divergent trace, we define the following renormalized
partition function in terms of a small kinetic term
$-\alpha {\left(\mu^{-1} d/d{\bf m}\right)}^2$:
\begin{equation}
\label{renopartition}
Z_{\rm ren}(\beta) = \lim_{\alpha \rightarrow 0+}
              {{\rm Tr}\left[ \exp\left(-{1\over 2}\alpha\hat{A}\right)
                       \hat{K}\exp\left(-{1\over 2}\alpha\hat{A}\right)
                       \right]
                \over
               {\rm Tr}\left[\exp\left(-\alpha\hat{A}\right)\right]
              },
\end{equation}
where $\hat{A}$ is the positive self-adjoint operator
$-{\left(\mu^{-1} d/d{\bf m}\right)}^2$
associated with the boundary condition that $\mu^{-1} d/d{\bf m}$ acting
on its eigenfunctions vanishes at the boundaries ${\bf m}=0$ and ${\bf
m}=2\lambda^2 B$ (see Appendix A in LW for more details).
Taking the limit
$\alpha \rightarrow 0+$ in (\ref{renopartition}) gives
\begin{eqnarray}
Z_{\rm ren}(\beta) &=&
{\left(
\int_0^{2\lambda^2 B} \mu d{\bf m}
\right)}^{-1}
\times
\nonumber
\\
&&\quad
\times
\int_0^{2\lambda^2 B} \mu d{\bf m}
\, \exp \left[ -\left(1-\sqrt{1- {\bf m} {(2\lambda^2 B)}^{-1}}
\right) 4\lambda^2 B \beta + 2\pi \hbar^{-1} \lambda^{-1}
{\bf m}\right].
\label{partition2}
\end{eqnarray}
In terms of the dimensionless variable $x = {(2\lambda^2 B)}^{-1}
{\bf m}$, (\ref{partition2})~can be written in the notation of
Ref.\cite{WYprl} as
\begin{equation}
\label{partition3}
Z_{\rm ren}(\beta) =
{\left(
\int_0^1 \mu dx
\right)}^{-1}
\int_0^1 \mu dx \exp \left[ - I_{*}(x)/\hbar \right] ,
  \end{equation}
where the effective action $I_{*}(x)$ is
\begin{equation}
\label{I}
I_{*}(x) = 4\pi \lambda B \left[ 2\left( 1 - \sqrt{1 - x} \right)
   {\beta \over \beta_c}- x \right],
  \end{equation}
and the critical inverse temperature $\beta_c$ is given by
\begin{equation}
\label{betac}
\beta_c = {2\pi \over \lambda \hbar}.
\end{equation}

\subsection{Thermodynamics}
\label{subsec:thermodynamics}

A~first observation from the partition function (\ref{partition3}) is that
the (constant volume) heat capacity, $C = \beta^2 \bigl( \partial^2 (\ln
Z_{\rm ren}) /
\partial \beta^2 \bigr)$, is always positive.\footnote{In general,
suppose that a partition function can be written as $Z(\beta) = \int dx \,
\nu(x) e^{-\beta x}$, where $\nu(x)\ge0$. As $Z (\partial^2 Z / \partial
\beta^2) - {(\partial Z / \partial \beta)}^2 =
\case{1}{2} \int dx\,dy \, {(x-y)}^2 \nu(x) \nu(y) e^{-\beta (x+y)}$, the
heat capacity, $C = \beta^2 \biglb( \partial^2 (\ln Z) /
\partial \beta^2 \bigrb)$, is then necessarily positive.}
The canonical ensemble with our boundary conditions is thus thermodynamically
stable. This is analogous to the stability of the fixed volume
canonical ensemble for spherically symmetric Einstein gravity in four
dimensions\cite{york1,WYprl}.

We shall now assume that $\mu$ varies slowly compared with the exponential
factor in~(\ref{partition3}). This will enable us to evaluate the integral by
the saddle point approximation.

The behavior of the partition function exhibits two qualitatively
different regions. Let us first suppose $\beta < \beta_c$. In this case
$I_{*}(x)$ has a global minimum at $x_0 = 1-{(\beta /
\beta_c)}^2$, and $I_{*}(x_{0}) = - 4 \pi \lambda B  {[ 1 - (\beta /
\beta_c) ]}^2<0$. The saddle point approximation yields
\begin{equation}
\label{partition}
Z_{\rm ren}(\beta) \sim
\exp \left[
- I_{*}(x_{0})/\hbar \right] .
\end{equation}
The dominant contribution therefore comes from a classical Euclidean black
hole solution with the mass ${\bf m}_0 = 2\lambda^2 B x_0$, whose Euclidean
action is~$I_{*}(x_{0})$. This is what one would have expected already
from the Lorentzian viewpoint: for the Lorentzian black hole solution, the
Hawking temperature at infinity is $\beta_c^{-1}$ for any value of the
mass\cite{Witten}, and (\ref{Rmetric}) shows that the local Hawking
temperature at a finite distance is obtained by multiplication with the
blueshift factor~$F^{-1/2}$. Note that for the dominating classical
solutions, $B$~can be arbitrarily large compared with the length scale
$\lambda^{-2}{\bf m}_0$ that is set by the mass. This feature is
qualitatively different from four-dimensional spherically symmetric
Einstein theory, where the mass of a dominant saddle point solution is
always so large that the box lies within the closed photon
orbit\cite{york1,WYprl}.
{}From (\ref{partition}) we recover for the energy expectation value
$\langle E \rangle$ the expression
\begin{equation}
\label{eenergy}
\langle E \rangle = - { \partial \left( \ln Z_{\rm ren} \right) \over
\partial \beta}
\simeq
4 \lambda^2 B \left[ 1 - (\beta/\beta_c) \right].
\end{equation}
Expressing $\beta$ in terms of ${\bf m}_0$ and inverting (\ref{eenergy})
yields
\begin{equation}
\label{mass-energy}
{\bf m}_0 \approx
\langle E \rangle
- { {\langle E \rangle}^2 \over 8 \lambda^2 B }.
\end{equation}
Eq.~(\ref{mass-energy}) gives an interpretation to the ADM mass as the
sum of the thermal energy and the gravitational self-energy associated
with the thermal energy. An analogous formula holds for the
four-dimensional Schwarzschild hole\cite{york1}.

The entropy associated with the gravitational field of a two-dimensional
black hole, $S_{GF}$, is given to the leading order by
\begin{equation}
\label{entropy}
S_{GF} = \left( 1 - \beta {\partial \over \partial \beta } \right)
\ln \biglb( Z_{\rm ren}(\beta) \bigrb)
\simeq
\beta_c {\bf m}_0.
\end{equation}
This is precisely the two-dimensional Bekenstein-Hawking black
hole entropy, $S_{BH} = \beta_c {\bf m}_0$\cite{Mathur}. The (constant
volume) heat capacity is
\begin{equation}
\label{heatcapacity}
C = \beta^2 { \partial^2 (\ln Z_{\rm ren}) \over
\partial \beta^2 }
\simeq
{4 \lambda^2 B \beta^2 \over \beta_c}.
\end{equation}

The higher order corrections to $\langle E \rangle$ and $S_{GF}$ depend
on the choice of the weight function~$\mu$. As an example, let us
consider the case where $\mu$ is independent of~${\bf m}$. To the
next-to-leading order, one then obtains
\begin{equation}
\label{next-entropy}
S_{GF}
\simeq
S_{BH}
- \case{1}{2}
\ln
\left(
\lambda B \over \hbar
\right)
+ \ln (\beta/\beta_c)
- 1.
\end{equation}
When the radius of the box is much larger than the horizon radius, we have
$\lambda B \gg \lambda^{-1} {\bf m}_0$ and $\beta/\beta_c \approx 1$. If in
addition $\lambda^{-1} {\bf m}_0 \gg \hbar$, so that the semiclassical
approximation is good, the next-to-leading order contributions to $S_{GF}$
are dominated by the second term on the right hand side
of~(\ref{next-entropy}).  Eq.~(\ref{next-entropy}) appears thus to be in
agreement with the quantum corrections from entanglement entropy discussed in
Refs.\cite{Mathur,Russo}. Note that the expression given in
(\ref{next-entropy}) does not involve a renormalization cutoff parameter;
however, a renormalization was involved in obtaining the
partition function from the divergent expression~(\ref{partition1}).

It is of interest to compare our partition function to the partition
function that is obtained from a Euclidean path integral via the
Hamiltonian reduction method that Whiting and York introduced in the
four-dimensional context\cite{WYprl,whitingCQG}. Adapting the
Whiting-York method to our case leads to a partition function that is
obtained from (\ref{partition2}) by replacing
${\left( \int \mu d{\bf m} \right)}^{-1} \mu d{\bf m}$ by $d(S_{BH}) = 2\pi
\lambda^{-1}\hbar^{-1} d{\bf m}$. We see that if $\mu$ is
chosen independent of ${\bf m}$, our partition function (\ref{partition3})
differs from the Whiting-York weighted partition function only by the
overall factor $4\pi\lambda B\hbar^{-1}$. The two partition functions thus
yield identical results for quantities that only involve logarithmic
temperature derivatives of the partition function, such as the energy
expectation value and the heat capacity.
The quantum corrections to the entropy differ, however: for the
Whiting-York weighting, (\ref{next-entropy})~is replaced by
\begin{equation}
\label{next-entropy-prime}
S_{GF}
\simeq  S_{BH}  + \case{1}{2}
\ln
\left(
\lambda B \over \hbar
\right)  + \ln (\beta/\beta_c)  - 1 + \ln(4\pi) .
\end{equation}

As $\beta$ approaches $\beta_c$ from below,
$\beta \rightarrow \beta_{c-}$, one has $x_0 \rightarrow 0$, and
${\lambda^2 B {\bf m}_0^{-1}}$ diverges. For fixed~$B$ this means that
the mass of the saddle point black hole approaches zero, and the saddle point
approximation is no longer expected to be good in this limit.
However, if one takes the limits $\beta \rightarrow \beta_{c-}$ and
$\lambda B\to\infty$ simultaneously, in such a way that $\lambda^{-1}{\bf
m}_0$ is fixed and much larger than~$\hbar$, the saddle point approximation
remains valid. In this limit, the energy expectation value $\langle E
\rangle$ becomes just the ADM mass, and the saddle point describes a black
hole of mass ${\bf m}_0$ in asymptotically flat space. Equivalently,
taking $\lambda B\to\infty$ while keeping $\lambda^{-1}{\bf
m}_0$ fixed implies, through the saddle point condition (for $\beta <
\beta_c$), that $\beta \rightarrow \beta_{c-}$. So, the temperature at
asymptotic infinity is $\beta_c$.
This is the
solution usually referred to as the Witten black hole.  Note that
in this limit, the (constant volume) heat capacity  given in
Eq.~(\ref{heatcapacity}) diverges. This is  consistent with the observation
that for the Witten black hole, the Hawking temperature at infinity is
independent of the mass: the hole can change its energy without changing the
temperature at infinity, and the heat capacity can thus be regarded as
infinite.

We finally turn to the case $\beta > \beta_c$. The global minimum of
$I_{*}(x)$ is now at $x=0$, and $I_{*}(x=0)=0$. There are no saddle points,
and the dominant contribution to (\ref{partition3}) comes from the vicinity
of $x=0$. Again, this agrees with Lorentzian expectations: for a Lorentzian
classical solution, the local Hawking temperature is always higher than the
Hawking temperature $\beta_c^{-1}$ at the infinity. As in four-dimensional
spherically symmetric Einstein gravity\cite{york1,WYprl}, one may see this
as evidence for a phase transition between a black hole sector and a
topologically different ``hot flat space" sector of the theory.
A~difference is, however, that in the dilatonic theory the transition is
sharply related to the existence of a saddle point. In the four-dimensional
case of Refs.\cite{WYprl,whitingCQG} the transition occurs while the
effective action still has a local minimum, but this minimum no longer
gives the dominant contribution to the partition function.

\section{Asymptotically flat space}
\label{sec:flat}

In this section we consider briefly the situation where the timelike
boundary is replaced by an asymptotically flat infinity.

Proceeding as in Ref.\cite{Madhavan}, one arrives at a classically reduced
action that can be obtained from Eqs.\ (\ref{classac}) and
(\ref{qh}) by taking the limit $\lambda B\to\infty$. $Q_B$~has then the
interpretation as the proper time elapsed at infinity. One can
quantize as in subsection~\ref{subsec:quantization}, the only difference
being that the range of ${\bf m}$ is now $0<{\bf m}<\infty$. The analytic
continuation of the time evolution operator is performed as in
subsection~\ref{subsec:partition}. The trace of the analytically
continued time evolution operator is again divergent, and the infinite
range of ${\bf m}$ renders the renormalization technique used in subsection
\ref{subsec:partition} not  directly applicable; however, in the limit
${\bf m} B^{-1}\rightarrow 0 $, the effective action in (\ref{I})
$I_* (x)$ approaches the expression $(\beta - \beta_c ) \hbar {\bf m}$, and
one can argue that the final
expression for the renormalized partition function
obtained from (\ref{partition3}) should be
\begin{equation}
Z_{\rm ren}(\beta) =
A \int_0^\infty \mu d{\bf m}
\, \exp \left[
- (\beta - \beta_c ) {\bf m}
\right],
\label{flat-partition}
\end{equation}
where the normalization factor $A$ depends on the details of the
renormalization, and may depend on $\lambda$, but is independent
of~$\beta$.

Let us again assume that $\mu$ varies slowly compared with the exponential
factor in~(\ref{flat-partition}). The expression (\ref{flat-partition})
is then divergent for $\beta<\beta_c$, but for $\beta>\beta_c$ the integral
converges and yields a well-defined partition function. This is in a
striking contrast with four-dimensional spherically symmetric Einstein
gravity, where the asymptotically flat space limit yields a divergent
integral for all values of the temperature\cite{WYprl,LW}. The reason for
this difference is that the second term $\beta_c {\bf m}$ in the exponent in
(\ref{flat-partition}) grows only linearly in ${\bf m}$, whereas in
four-dimensional spherically symmetric Einstein gravity the corresponding
term grows quadratically in the mass variable. Note that in both cases
this term can be interpreted as the Bekenstein-Hawking entropy\cite{WYprl}.

We now assume $\beta>\beta_c$. The heat capacity is again positive, and
the canonical ensemble is thus stable. However, the integral in
(\ref{flat-partition}) does not admit a saddle point approximation, and the
partition function gets its dominant contribution from the vicinity of
${\bf m}=0$. This is analogous to what happened also in the finite
boundary case for $\beta>\beta_c$, and reflects the fact that there
are no classical black hole solutions with the Hawking temperature at
infinity lower than~$\beta_c^{-1}$.

For concreteness, let us set $\mu = {\bf m}^p$, $p>-1$ in
(\ref{flat-partition}). The energy
expectation value, the heat capacity, and the entropy are then given by
\begin{mathletters}
\begin{eqnarray}
\langle E \rangle
&=&
{(p+1) \over (\beta - \beta_c)},
\label{eenergy-p}
\\
C
&=&
{(p+1) \beta^2 \over {(\beta - \beta_c)}^2 },
\label{heatcap-p}
\\
S &=&
(p+1)
\left[
{\beta_c \over \beta - \beta_c}
+ \ln \left( \beta_c \over \beta - \beta_c \right)
\right]
\; + \; \hbox{constant}.
\end{eqnarray}
\end{mathletters}%
At the limit $\beta\to\beta_{c+}$, both $\langle E \rangle$, $C$, and $S$
diverge. A~way to understand the divergence in $\langle E \rangle$ physically
is to recall that a classical two-dimensional black hole solution satisfies
$\beta=\beta_c$ with any value of the mass. For $\beta\to\beta_{c+}$,
arbitrarily high mass black holes would thus be expected to contribute to
$\langle E \rangle$ with roughly equal weights, resulting in a divergence.
A~similar interpretation accounts for the divergence in~$S$. The divergence
in $C$ can be interpreted as in subsection~\ref{subsec:thermodynamics}, in
terms of the fact that the Hawking temperature at infinity is independent of
the mass.

\section{Concluding remarks}
\label{sec:conclusions}

In this paper we have investigated the equilibrium thermodynamics of the
two-dimensional vacuum dilatonic black hole in the canonical ensemble. The
classically reduced Hamiltonian theory of Ref.\cite{Lau} was quantized, and
the thermodynamical partition function was obtained as the trace of the
analytically continued time evolution operator. When the system is confined
in a finite box that is characterized by the boundary value of the dilaton
field, the partition function is well-defined for all boundary temperatures,
and the heat capacity is always positive. For temperatures higher than
$\beta_c^{-1} = \hbar\lambda/(2\pi)$, the partition function is dominated
by a classical black hole solution, and the dominant contribution to the
entropy is the two-dimensional Bekenstein-Hawking entropy, $S_{BH}=\beta_c
{\bf m}_0$, where ${\bf m}_0$ is the mass of the hole. The situation is thus
qualitatively very similar to that with four-dimensional
Schwarzschild black holes\cite{york1,WYprl,LW}. The main
difference is that in our two-dimensional case the condition for a saddle
point to dominate the partition function only depends on the temperature,
whereas in the Schwarzschild case the corresponding condition involves both
the temperature and the boundary radius.

In the limit of asymptotically flat boundary conditions our partition
function remains well-defined for temperatures lower than~$\beta_c^{-1}$.
The heat capacity is again positive, but the partition function cannot be
approximated by a classical black hole solution. When the temperature
approaches $\beta_c^{-1}$, the energy expectation value, the heat capacity,
and the entropy all diverge, provided the measure in the partition function
is sufficiently slowly varying in the mass
variable; this divergence can be understood
physically in terms of the fact that for a classical black hole, the Hawking
temperature at infinity is independent of the black hole mass. This behavior
is in a striking contrast with four-dimensional spherically symmetric
Einstein gravity, where the asymptotically flat space canonical ensemble
does not exist for any temperature\cite{york1,WYprl}. The underlying reason
for the difference is that for the dilatonic black hole the Bekenstein-Hawking
entropy $S_{BH}$ is linear in the mass, whereas for the four-dimensional
Schwarzschild black hole $S_{BH}$ is quadratic in the mass.

When the partition function of the finite boundary canonical ensemble is
dominated by a classical black hole solution, the next-to-leading order
corrections to the energy expectation value and to the entropy depend
sensitively on the inner product that is adopted in the Lorentzian
Hamiltonian quantum theory. If the weight factor $\mu$
in the inner product is chosen to
be independent of the mass variable, the next-to-leading order correction to the
entropy appears to be compatible with the first quantum corrections from the
entanglement entropy\cite{Mathur,Russo}. The correction does not involve an
explicit renormalization parameter; however, a renormalization was required
when recovering the partition function from a formally divergent trace.
These results seem to be in agreement with Frolov's observation
\cite{Frolov} that the 1-loop {\em thermodynamical\/} entropy is finite,
while the {\em entanglement\/} (or statistical mechanics) 1-loop entropy
diverges and requires a renormalization cutoff\cite{Srednicki}.
The study of two-dimensional
black holes in the presence of fields other than the  gravitational field
may help to clarify this point\cite{Romano}.

It might be interesting to investigate along the lines of the present paper
the thermodynamics of the one-loop corrected model of Ref.\cite{BPP} with
matter fields. Work in this direction is in progress.

\acknowledgments
We would like to thank Karel Kucha\v{r}, Joe Romano, and Madhavan
Varadarajan for stimulating discussions. We
would also like to thank John Friedman and Bernard Whiting for helpful
discussions and for their constructive criticism of the manuscript. This
work was supported in part by NSF grants PHY-91-19726 and PHY-95-07740.


\begin{references}

\bibitem{Witten}
E.~Witten,
Phys.\ Rev.\ D {\bf44}, 314 (1991).

\bibitem{CGHS}
C.~G. Callan,
S.~Giddings,
J.~Harvey, and
A.~Strominger,
Phys.\ Rev.\ D {\bf45}, 1005 (1992).

\bibitem{stro-rev}
A.~Strominger,
1994 Les Houches Lectures on Black Holes,
hep-th/9501071.

\bibitem{GH1}
G.~W. Gibbons and
S.~W. Hawking,
Phys.\ Rev.\ D {\bf15}, 2752 (1977).

\bibitem{hawkingCC}
S.~W. Hawking, in
{\it General Relativity: An Einstein
Centenary Survey,}
edited by S.~W. Hawking and W.~Israel
(Cambridge University Press, Cambridge, 1979).

\bibitem{york1}
J.~W. York,
Phys.\ Rev.\ D {\bf 33}, 2092 (1986).

\bibitem{WYprl}
B.~F. Whiting and
J.~W. York,
 Phys.\ Rev.\ Lett.\ {\bf 61}, 1336 (1988).

\bibitem{whitingCQG}
B.~F. Whiting,
Class.\ Quantum Grav.\ {\bf 7}, 15 (1990).

\bibitem{pagerev}
D.~N. Page, in
{\it Black Hole Physics,}
edited by V.~D. Sabbata and Z.~Zhang
(Kluwer Academic Publishers, Dordrecht, 1992).

\bibitem{LWo}
J.~Louko and
B.~F. Whiting,
Class.\ Quantum Grav.\ {\bf 9},
457 (1992).

\bibitem{BY-quasilocal}
J.~D. Brown and
J.~W. York,
Phys.\ Rev.\ D {\bf47}, 1407 (1993).

\bibitem{BY-microcan}
J.~D. Brown and
J.~W. York,
Phys.\ Rev.\ D {\bf 47},
1420 (1993).

\bibitem{MW}
J.~Melmed and
B.~F. Whiting,
Phys.\ Rev.\ D {\bf 49}, 907
(1994).

\bibitem{LW}
J. Louko and
B.~F. Whiting,
Phys.\ Rev.\ D {\bf51}, 5583
(1995).

\bibitem{Solodukhin}
S.~N.~Solodukhin, ``Two-dimensional
Quantum-Corrected Eternal Black Hole'',
hep-th/9506206.

\bibitem{KVK}
K.~V. Kucha\v{r},
Phys.\ Rev.\ D {\bf50}, 3961
(1994).

\bibitem{Madhavan}
M.~Varadarajan, ``Classical and quantum
geometrodynamics of 2d vacuum dilatonic black holes",
gr-qc/9508039.

\bibitem{Lau}
S.~R. Lau, ``On the canonical reduction of spherically
symmetric gravity",
Report TUW-95-21, gr-qc/9508028.

\bibitem{bilal}
A.~Bilal and
I.~I. Kogan,
Phys.\ Rev.\ D {\bf 47}, 5408 (1993).

\bibitem{Horowitz}
G.~W. Gibbons and K.~Maeda,
Nucl.\ Phys.\ {\bf B298},
741 (1988);
D.~Garfinkle,
G.~T. Horowitz, and
A.~Strominger,
Phys.\ Rev.\ D {\bf 43}, 3140 (1991);
S.~B. Giddings and
A.~Strominger,
Phys.\ Rev.\ D {\bf 46}, 627 (1992).

\bibitem{Peleg}
Y.~Peleg,
Mod.\ Phys.\ Lett.\ {\bf A9}, 3137 (1994).

\bibitem{ADM}
R.~Arnowitt,
S.~Deser, and
 C.~W. Misner, in
{\it Gravitation: An Introduction to Current Research},
ed.\
L.~Witten (Wiley, New York, 1962);
T.~Regge and
C.~Teitelboim,
Ann.\ Phys.\ (N.Y.), {\bf 88}, 286 (1974).

\bibitem{Mathur}
E.~Keski-Vakkuri and
S.~Mathur,
Phys.\ Rev.\ D {\bf50},  917 (1994);
T.~Fiola,
J.~Preskill,
A.~Strominger, and
S.~Trivedi,
Phys.\ Rev.\ D {\bf50}, 3987 (1994);
R.~C. Myers,
Phys.\ Rev.\ D {\bf50},  6412 (1994).

\bibitem{Russo}
J. Russo,
``Entropy and black hole horizons", CERN report
no. CERN-TH/95-179, hep-th/9507009 (1995).

\bibitem{Frolov}
V.~P. Frolov,
Phys.\ Rev.\ Lett.\ {\bf 74}, 3319 (1995).

\bibitem{Srednicki}
L.~Bombelli,
R.~Koul,
J.~Lee, and
R.~D.~Sorkin,
Phys.\ Rev.\ D {\bf 34}, 373 (1986);
M.~Srednicki,
Phys.\ Rev.\ Lett.\ {\bf 71}, 666 (1993);
G.~'t Hooft,
Nucl.\ Phys.\ {\bf B256}, 727 (1985);
V.~Frolov and
I.~Novikov,
Phys.\ Rev.\ D {\bf48},  4545 (1993);
L.~Susskind and
J. Uglum,
Phys.\ Rev.\ D {\bf50},  2700 (1994);
M. Maggiore,
Nucl.\ Phys.\ {\bf B429}, 205 (1994);
C.~Callan and
F. Wilczek,
Phys.\ Lett.\ {\bf B333}, 55 (1994);
S. Carlip and
C. Teitelboim,
Phys.\ Rev.\ D {\bf51}, 622 (1995);
S. Carlip,
Phys.\ Rev.\ D {\bf51}, 632 (1995);
Y.~Peleg,
``Quantum Dust Black Holes'', Brandeis University report no.\
BRX-TH-350, hep-th/93077057 (1993).

\bibitem{Romano}
J.~Romano,
private communication.

\bibitem{BPP}
S.~Bose,
L.~Parker, and
Y.~Peleg,
Phys.\ Rev.\ D {\bf 52},  3512 (1995);
Phys.\ Rev.\ Lett. {\bf 76}, 861 (1996).

\end{references}
\end{document}